\documentclass[12pt]{article}

\usepackage[dvips]{graphicx}
\def\leq{\underline{<}}
\begin{document}
\title{
Packing and percolation of poly-disperse discs and spheres
}
\author
{Takashi Odagaki, Tsuyoshi Okubo, 
Ryusei Ogata\footnote{Present address: NEC Corporation, Tokyo, Japan},\\
and Keiji Okazaki\dag\\
Department of Physics, Kyushu University, Fukuoka 812-8581 Japan\\
\hspace{-0.3cm}\dag\ Mitsubishi Chemical Corporation,\\ Science and
Technology Research Center, Yokkaichi 510-8530 Japan}

\maketitle
\begin{abstract}
Random dense packing of poly-disperse discs and spheres generated
under the infinitesimal gravity protocol is investigated by a Monte Carlo
simulation.
For the binary discs packed in two dimensions, the packing fraction of disc
assembly becomes lower than that of the monodisperse system
when the size ratio is close to unity. 
We show that the suppressed packing fraction is
caused by an increase of the adjacent neighbours with long bonds
where the adjacent neighbours is defined on the basis of
the Laguerre (radical) tessellation. 

For the poly-disperse systems in two and three dimensions,
the packing fraction is shown to have a minimuma as a function
of the poly-dispersity.
Percolation process in the densely packed discs and spheres is also studied.
The critical area (volume) fraction in two (three) dimensions is shown to be
a monotonically increasing (decreasing) function of the poly-dispersity.
\end{abstract}


\section{Introduction}

Random close packing of spheres is an important concept in understanding
the structure and properties of liquids and glasses.
The first significant contribution in this direction was
reported by Bernal\cite{Bernal} who analyzed the structure of glasses
on the basis of the sphere packing.
It is also well known that solidification of liquids\cite{Alder} and
the flow of granular materials\cite{Jaeger} are closely related to
packing properties of hard spheres.

In practical applications, it is common to encounter systems in which the
size of dispersed particles is not unique but has a certain distribution.
For example,  a functional composite system
can be fabricated by depositing nano-clusters which are synthesized by the
plasma-gas-condensation technique.
In this case, the size distribution of the
nano-clusters plays important role in determining
the properties of the composite system\cite{katoh}.
Beside the packing fraction, the connectivity of particles
is also important characteristics of the system composed with dispersed
spheres. For example, 
a printing ink consists of carbon blacks dispersed in a varnish,
and the network formation of carbon blacks is one of the key parameters
controlling the properties of the ink. 
It is, therefore, important in practical applications
to find out the dependence of packing fraction and connectivity
on the width of the distribution or the poly-dispersity.

In this paper, we investigate the packing fraction for a random dense packing
of binary hard discs and of poly-disperse hard discs and spheres. 
In order to avoid subtlety in the definition of random close
packing\cite{MJS1},
we study the dense packing produced by the drop-and-roll or
the infinitesimal gravity protocol using a Monte Carlo simulation.
Although the structure produced by this protocol is not the random close
packing nor the maximally random jammed state\cite{MJS1,MJS2},
it is worth studying the structure produced by this sequential
protocol since in practical applications the packing of particles is
produced by a fixed procedure such as molecular vapour deposition or
dispersing particles in a solvent.
In \S 2 we explain briefly the protocol employed.
The packing of binary discs in two dimensions is studied in \S 3, where
results of computer simulation and an analysis based on the Laguerre
tessellation are given.
In \S 4, we present results for the packing of poly-disperse
discs in two dimensions and spheres in three dimensions.
For the poly-disperse systems we also study the percolation process
of randomly selected discs and spheres in the packed structure.
Conclusion is given in \S 5.

\section{Infinitesimal gravity protocol}
In the infinitesimal gravity  protocol\cite{okubo},
a container of $L \times L$
($L \times L \times L$) is prepared in two (three) dimensional space.
A particle (a disc in two dimensions and a sphere in three dimensions)
is introduced far above the container at a horizontal
position selected randomly. The radius
of the particle is chosen so that the distribution of the radius $R$ obeys
a given function $\phi(R)$.
Then, the particle is dropped vertically towards the bottom of the
container. The particle drops freely or rolls down around a
particle that it touches until the particle settles into a stable position
or makes contact with the base line. (See Fig.1.)
\begin{figure}[ht]
\begin{center}
\includegraphics[height=7cm,clip]{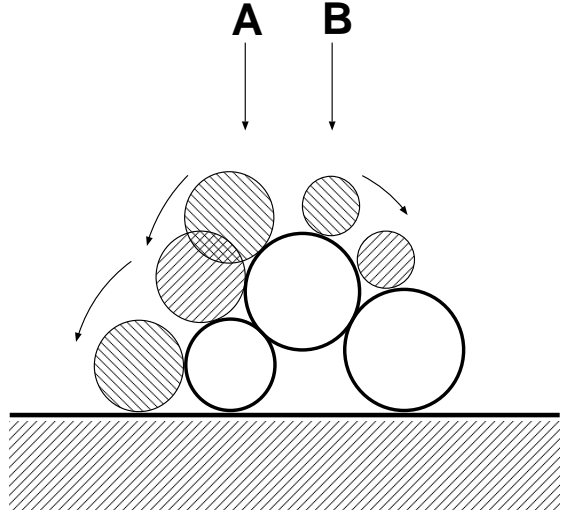}
\end{center}
\caption{The infinitesimal gravity protocol is illustrated schematically.
After contacting with another disc, each disc, A and B, rolls down
around the target disc maintaining contact until it makes contact with another
disc. It rolls down around the new disc if it can (A), and if it cannot roll 
down further, then it settles there (B).
}
\end{figure}
We repeat the process until the container is filled with particles.
It is known that this protocol produces for a mono-disperse system
a structure very close to the random close packing, though the structure is not
the maximally jammed one\cite{MJS1,MJS2}.

\section{Binary discs}
\subsection{Monte Carlo simulation}
Using the infinitesimal gravity protocol, we produced packed structures of
two kinds of discs, one with radius $R_S$ and the other with radius $R_L$.
The system size is set to $L/2R_L = 100$
and discs at given fraction $x$ of the smaller disc are packed up to a height
of $140 \times 2R_L$. To reduce the effects of
the boundaries, upper and lower $20 \times 2R_L$ layers are discarded
(so that the remaining area is $100 \times 100 \times 4R_L^2$), and we imposed
a periodic boundary condition in the horizontal direction.
Then, the packing fraction $f$ is obtained as a function of the disc size
ratio $r = R_S/R_L$ and the area fraction of smaller discs $n = x R_S^2
/[(1-x)R_L^2 + x R_S^2]$.

The packing fractions $f(n, r )$ obtained from the Monte Carlo simulation
are illustrated in Fig. 2.
\begin{figure}[ht]
\begin{center}
\includegraphics[height=7cm,clip]{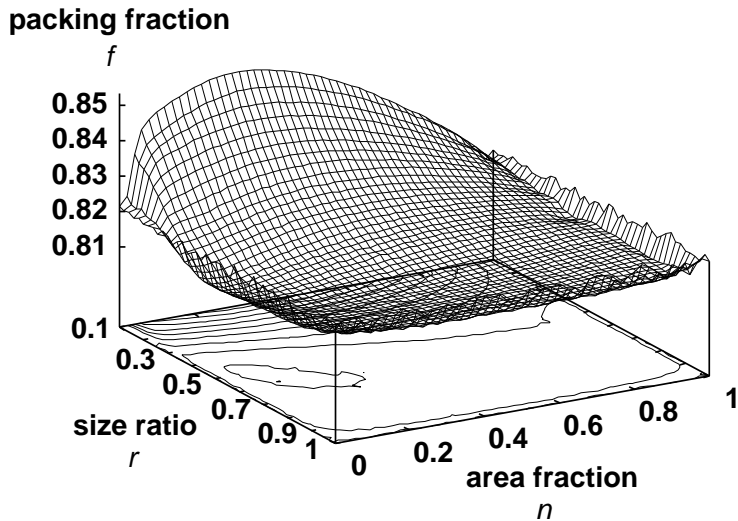}
\end{center}
\caption{A 3D plot of the packing fraction $f$ of a binary hard disc assembly
as a function of size ratio $r$ and area fraction $n$.
Each value was determined from an average of 20 samples. The contours
are drawn from 0.81 to 0.85 every 0.05.
}
\end{figure}
Note that $f(0, r) = f(1, r ) = f(n, 1) = f(n, 0) = f_2 \simeq 0.82$.
Here, $f_2$ is the packing fraction of the random close packing of monodisperse
discs\cite{berryman}.
For $r \simeq 0$, as $n$ is increased from $n = 0$, $f(n, r )$ increases
dramatically, because of the vacant spaces
between larger discs being filled by smaller discs.
The increase of packing fraction reaches a
maximum at a certain $n$ where the vacant spaces are almost filled up,
and decreases again to the value for the monodisperse assembly.
On the other hand, $f(n, r )$ for $r \simeq 1$
decreases from $f_2$ as $n$ is increased from $n = 0$, and is almost
constant over the wide range of $n$ until it increases again
to $f_2$ near $n = 1$.
The decrease of $f(n, r )$ is due to the fact that the
smaller disc, which is too large to fill the vacant spaces, replaces the
larger discs and destroys the structure formed by larger discs\cite{okubo}.

It is interesting to compare the contour for the packing fraction with the
contour of the constant mean squared deviation of radii scaled by the mean
diameter,
$(\langle R^2 \rangle - \langle R \rangle^2)/4\langle R \rangle^2$,
in the $(n, r)$ plane, which is shown in Fig. 3.
Here, $\langle R \rangle$ and $\langle R^2 \rangle$ are the mean and the mean
squared radius of discs in the system, respectively.
\begin{figure}[ht]
\begin{center}
\includegraphics[height=7cm,clip]{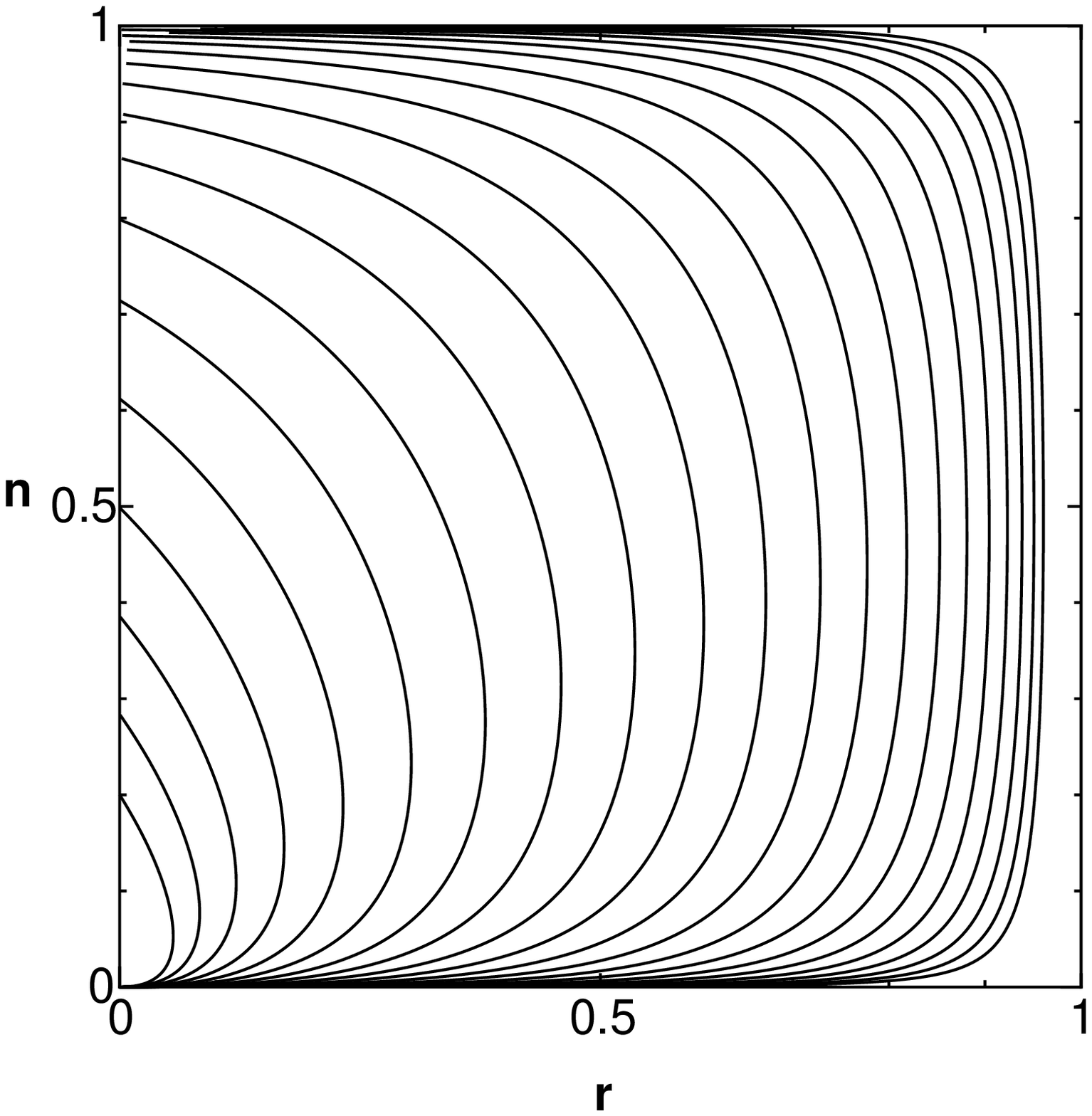}
\end{center}
\vspace{-1cm}
\caption{Contour plot of the constant mean squared deviation of radii
of discs. Contours are shown for $\log \sigma = -2 \sim 0$ for every
$0.1$, where $\sigma\equiv \sqrt{\langle R^2 \rangle -
\langle R \rangle^2}/2\langle R \rangle$.
}
\end{figure}
The similarity of these contours
suggests that the packing fraction is determined mostly by the
relative width of the distribution function.
This observation is strongly supported
by the results for poly-disperse systems discussed in \S 4.

\subsection{Analysis by the Laguerre tessellation}
Using the Laguerre tessellation\cite{Laguerre},
we can define adjacent neighbours of a disc and connect centres
of all adjacent neighbour pairs to form a triangular network
which fills the space without gaps.

For an infinite system, the mean number of adjacent neighbours for each disc
is equal to 6, which is a consequence of Euler's condition, and
therefore the number of triangles included
in the system is twice as many as the number of discs.
As a result, the packing fraction of the
system $f$ is expressed in term of the mean area of triangles $A_m$ as
\begin{equation}
f = \frac{\pi \langle R^2\rangle}{2A_m}.
\end{equation}

We analysed the bond lengths in detail. The bond length distribution in our
simulation is shown in Fig. 4 for the bond between two adjacentlarge
discs. It is apparent from Fig. 4 that
the major difference of bond distributions is in the
behaviour of the long bond region.
We can conclude that, although the tail is small compared with the whole
distribution, the behaviour of the tail greatly affects the packing
fraction\cite{okubo}.
\begin{figure}[ht]
\begin{center}
\includegraphics[height=7cm,clip]{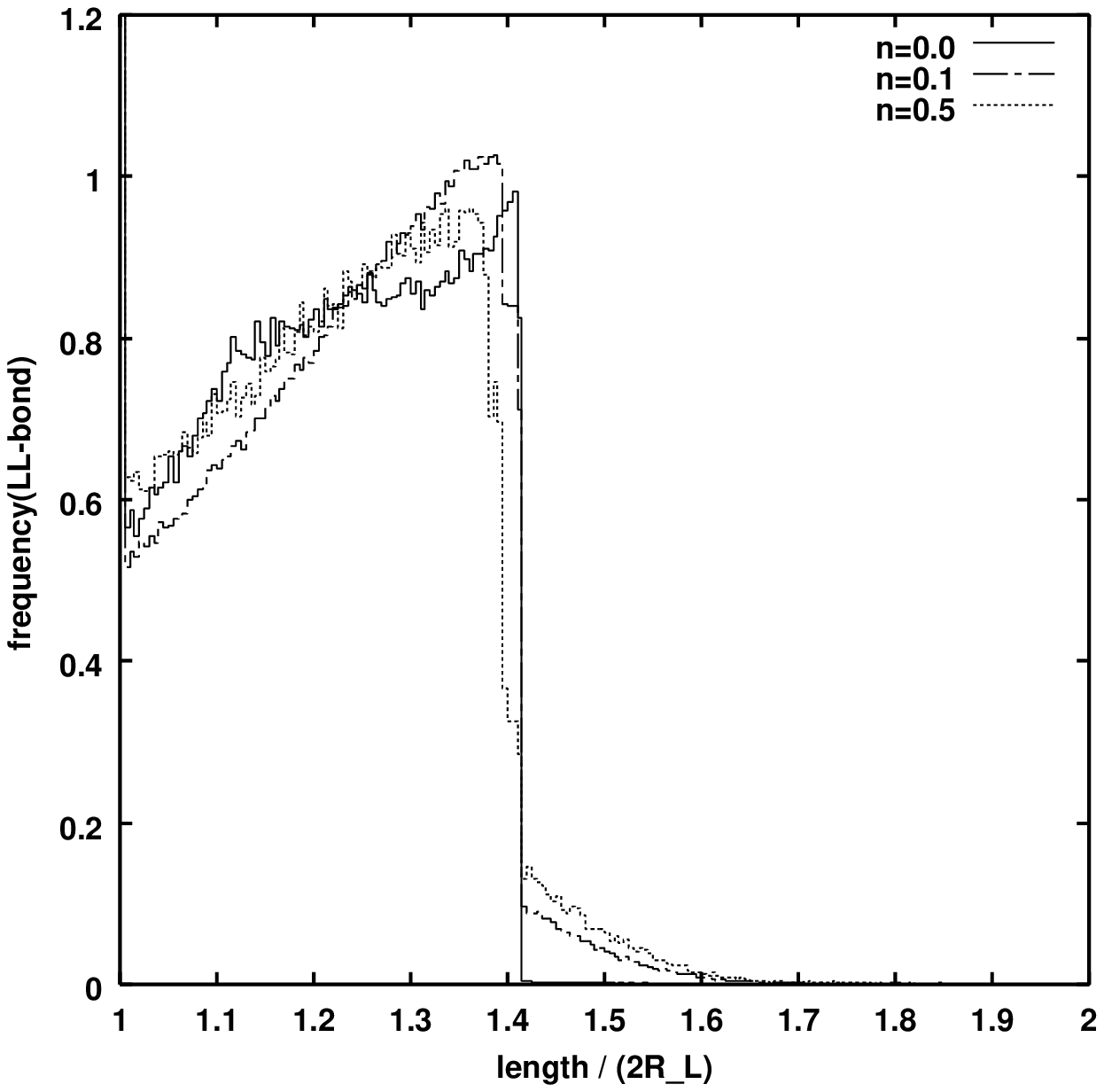} 
\end{center}
\caption{The distribution function of bond length between two large discs
at $r = 0.9$.
The solid, dash-dot and dot curves correspond to $n = 0.0$ (the
mono-diserse assembly), $n = 0.1$ and $n = 0.5$. Note that each
distribution function has a delta function-like peak at the length
which equals $2R_L$.}
\end{figure}

\section{Poly-disperse discs and spheres}
\subsection{Packing fraction}
We examined the Gaussian and the uniform distributions
for two and three dimensions. For two dimensions, we also examined
the log-normal distribution.
All distribution functions are normalized so that the average
radius $\int R \phi(R)dR$ is equal to $m$
and the mean squared deviation $\int (R-m)^2 \phi(R) dR$ is equal
to $\sigma^2$.
We set the average diameter $2m$ as the unit of length and use
the width $\sigma$ as a parameter representing the poly-dispersity.

We prepared a $200 \times 200$ ($60 \times 60 \times 60$) square (cubic) box
and examined 1000 samples in two dimensions and 3000 samples
in three dimensions to determined the packing fraction.
In this simulation, fixed boundary conditions are imposed
in horizontal directions.

Figures 5 and 6 show the dependence of $f$ on the poly-dispersity $\sigma$.
\begin{figure}[ht]
\begin{center}
\includegraphics[height=7cm,clip]{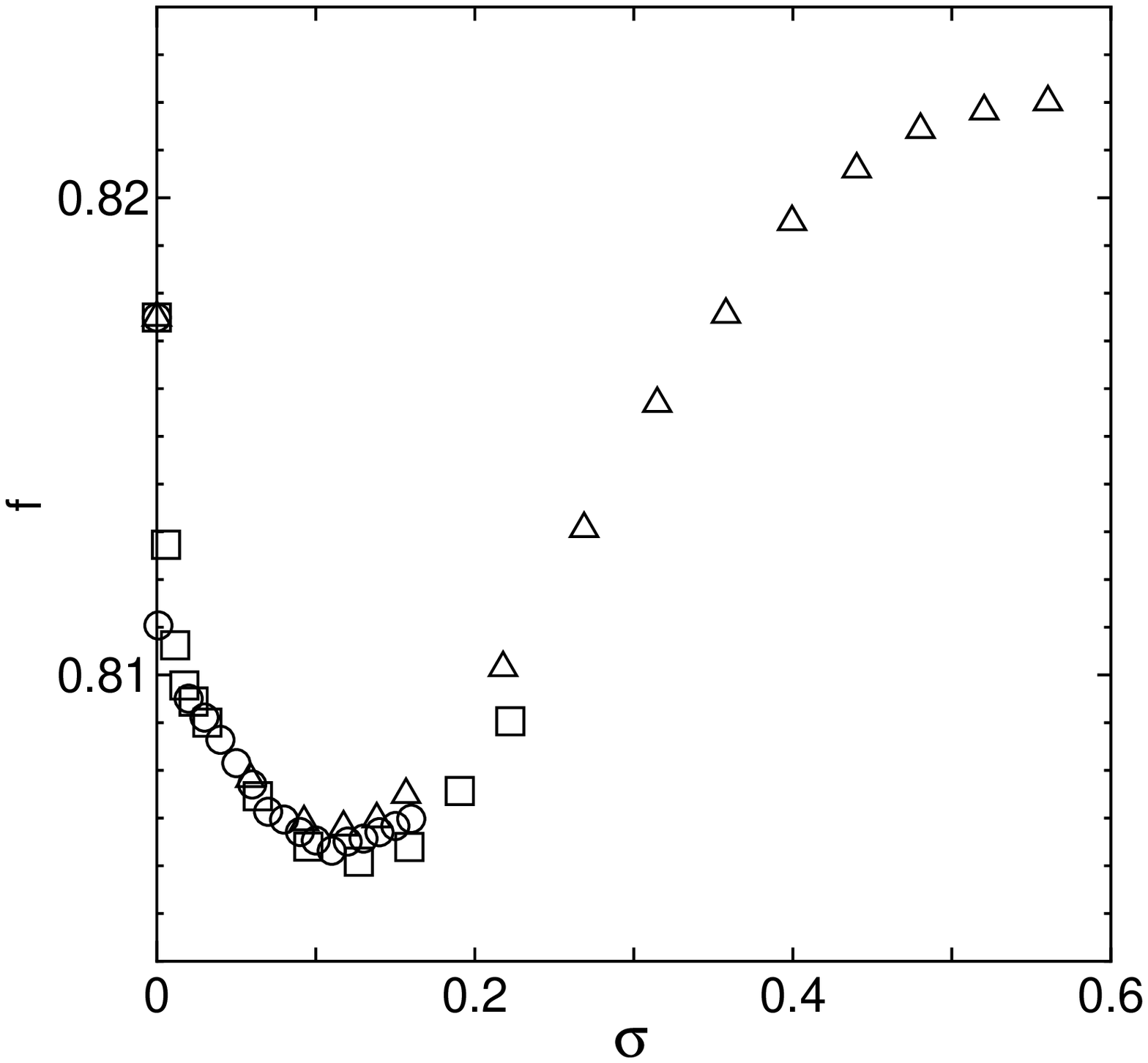} 
\end{center}
\caption{The relation between $f$ and $\sigma$ for various size distribution
functions in two dimensions: The results are shown 
for Gaussian (circles), uniform (squares),
and log-normal  (triangles) distributions.}
\end{figure}
The packing fraction $f=0.8175\pm0.0002$ at $\sigma = 0$ in two dimensions
is consistent with  the value $f_2 = 0.82$ reported in 
literature\cite{berryman}. Similarly, $f=0.5575\pm0.0003$
at $\sigma = 0$ for three dimensions agrees well with the
value $f_3 = 0.555 \pm 0.005$ reported by Onoda {\it et al.}\cite{Ono}. 
When $\sigma$ is increased, the packing fraction $f$ decreases
and the dependence is roughly independent of
the distribution function. When $\sigma$ is increased further,
the packing fraction $f$ exhibits a minimum and increases again.
This behaviour is identical to the behaviour of the packing fraction
of binary discs discussed in \S 3.
\begin{figure}[htb]
\begin{center}\includegraphics[height=7cm,clip]{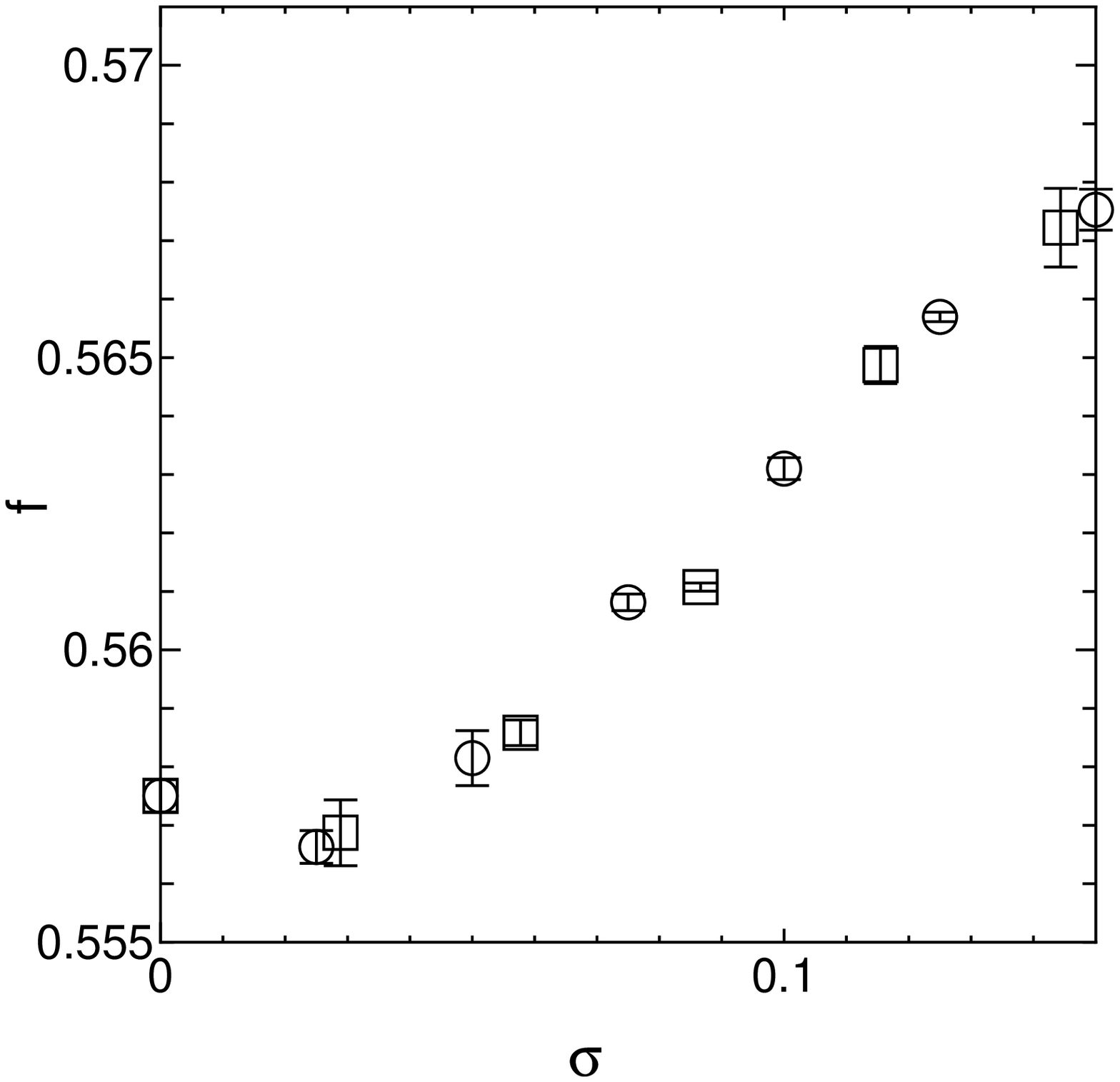} 
\end{center}
\caption{The relation between $f$ and $\sigma$ for two size distribution
functions in three dimensions: The results are shown for Gaussian (circles)
and uniform (squares) distributions.}
\end{figure}

\subsection{Continuum percolation}
In order to construct percolation process in the random assembly
of poly-disperse discs and spheres,
we randomly select a given fraction of the total number of
particles (coloured white at the beginning), irrespective of their size,
and change their colour to black. Two black particles are
regarded as connected when they touch each other.
Then, we investigate the percolation of the black particles
when their fraction is increased.

We determined the critical percolation area (volume) fraction  $A_c$  ($V_c$)
of black particles as a function of $\sigma$, where percolation
was judged by the appearence of a cluster spanning from the
bottom to the top surface. 
In Figs. 7 and 8, $A_c$ and $V_c$ are plotted as a function of $\sigma$.
Our value $A_c=0.4777\pm0.0005$ and $V_c=0.1938 \pm 0.0001$ at $\sigma=0$
are close to the critical area fraction determined from
lattice models $A_c=0.45\pm0.02$\cite{scher} and the value
of $V_c=0.185 \pm 0.005$ in literature\cite{Bou}.
\begin{figure}[ht]
\begin{center}
\includegraphics[height=7cm,clip]{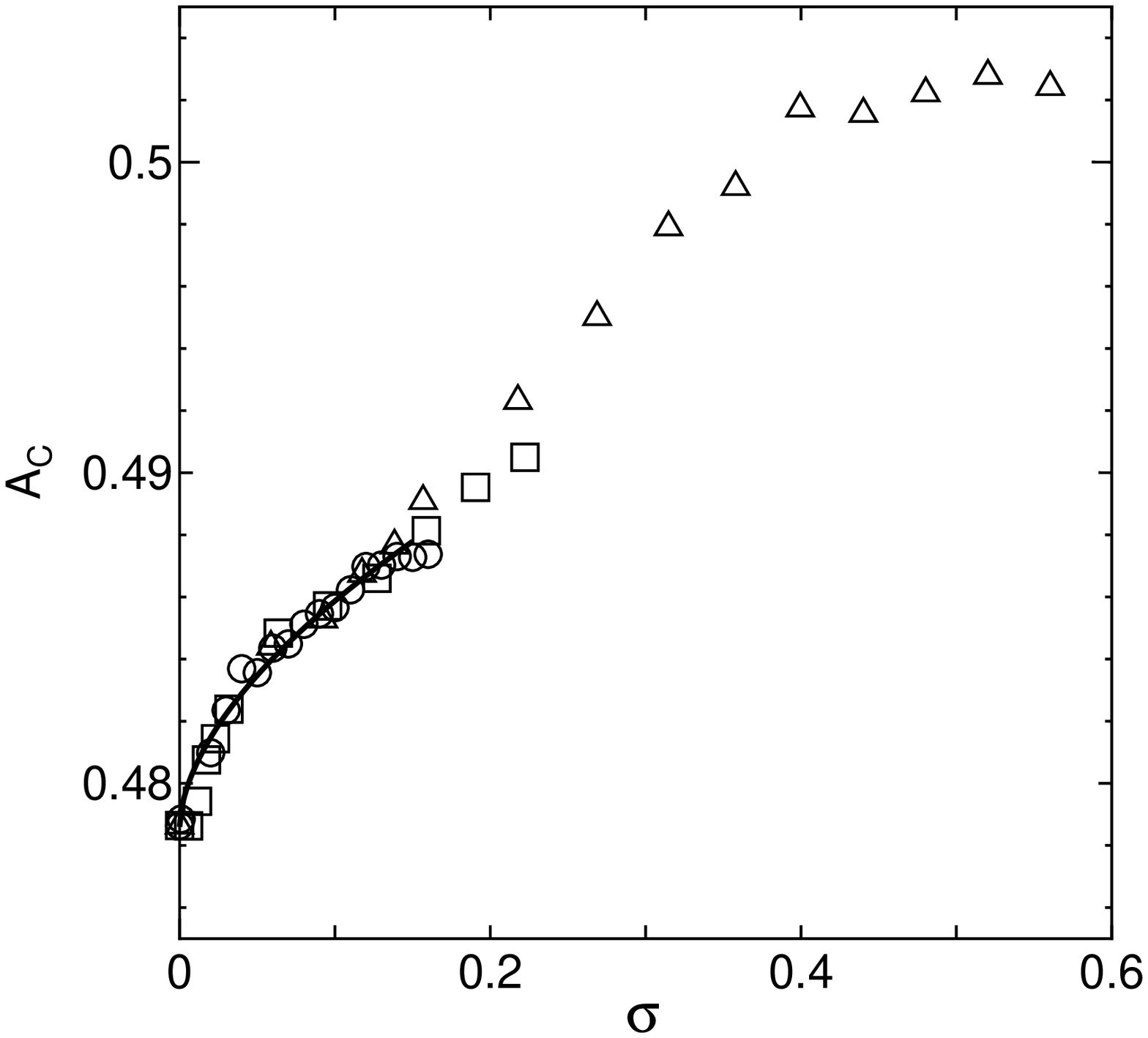} 
\end{center}
\caption{The critical area fraction $A_c$ is plotted against $\sigma$ for
three size distribution functions: Gaussian (circles), 
uniform (squares) and log-normal  (triangles). The solid line represents
the fitting function (\ref{ac-eq}).
Error bars are much smaller than the size of the symbols.
}
\end{figure}

The critical area fraction  $A_c$ is an increasing function of $\sigma$ and
is roughly independent of the distribution function when $\sigma$ is small.
In fact, if we use a power-law function
\begin{equation}
A_c(\sigma)=a_2 + b_2 \sigma^{c_2}
\label{ac-eq}
\end{equation}
to fit the data for $0 \leq \sigma \leq 0.15$, parameters are give by
\begin{equation}
a_2 = 0.4777\pm 0.0005, \quad b_2= 0.028 \pm 0.002, \quad c_2= 0.54 \pm 0.05.
\end{equation}
The solid line in Fig. 7 represents Eq. (\ref{ac-eq}).

The critical  volume fraction  $V_c$ is a decreasing function of $\sigma$.
Although the dependence is weak, this is an amazing contrast to the
critical area fraction $A_c$ in two dimensions. For $0 \leq \sigma \leq 0.15$,
$V_c$ can well be fitted by the following line
\begin{equation}
V_c(\sigma) =  a_3 + b_3 \sigma
\label{vc-eq}
\end{equation}
with
\begin{equation}
a_3 = 0.1938 \pm 0.0001, \qquad b_3= -0.005 \pm 0.001,
\end{equation}
which is shown by the solid line in Fig. 8.
\begin{figure}[ht]
\begin{center}\includegraphics[height=7cm,clip]{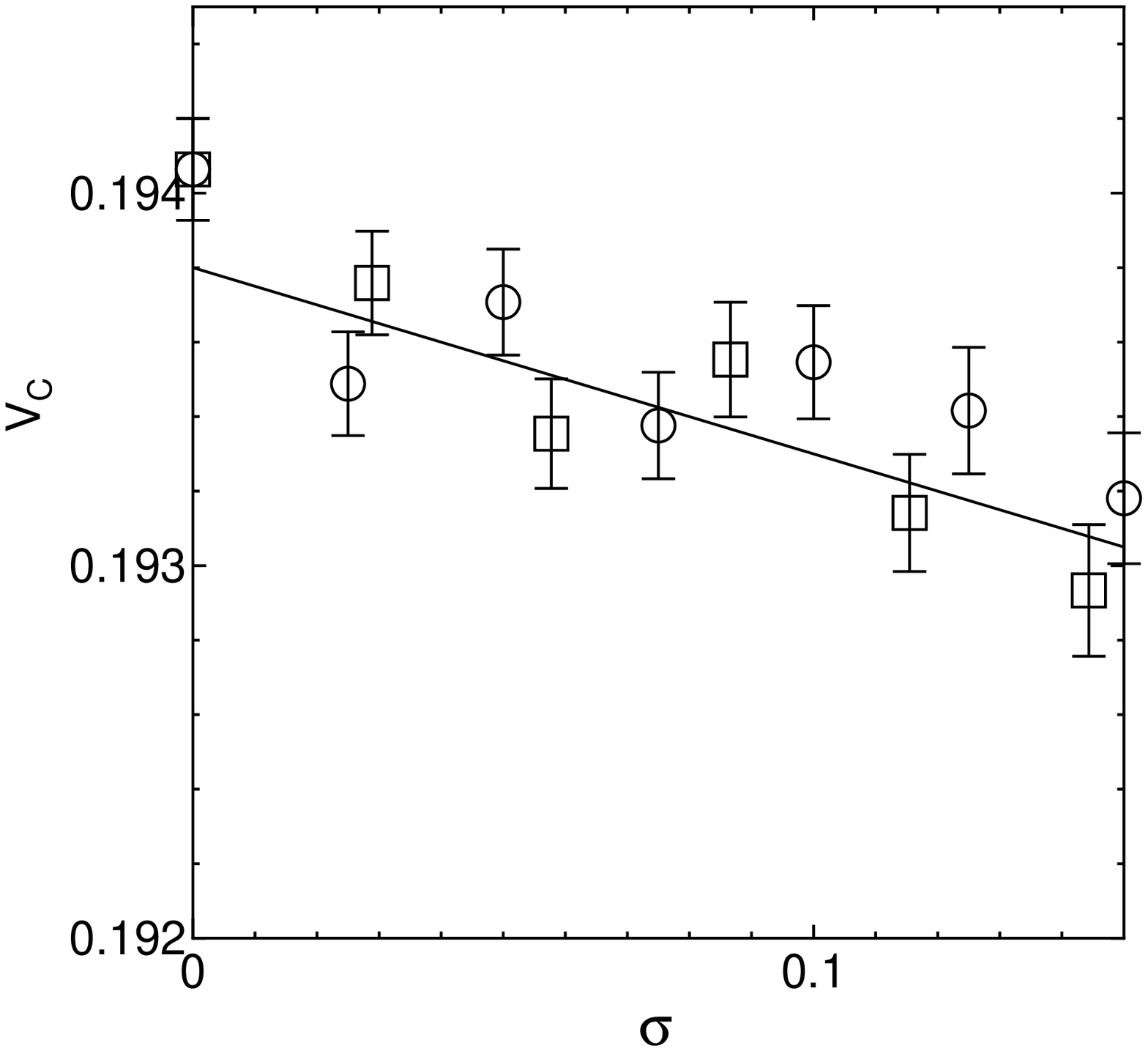} 
\end{center}
\caption{The critical volume fraction $V_c$ is plotted against $\sigma$ for
two size distribution functions:  Gaussian (circles) and
uniform (squares). The solid line represents function (\ref{vc-eq}).}
\end{figure}

The significant difference in the poly-dispersity dependence of the critical
area and volume fractions can be explained by the
surface area of discs and spheres which determines the effectiveness of
connection\cite{Ogata}. In fact, we can easily show that while the average
circumference of discs is independent of $\sigma$,
the average surface area of spheres increases as $\sigma$ is increased.

\section{Concluding remarks}
We have examined by Monte Carlo simulation the packing fraction of
poly-disperse discs and spheres which are packed in two and three dimensions
by the infinitesimal gravity protocol.
It is a striking result that the packing fraction decreases
for the weak poly-dispersity. This is due to the fact that a little
increase of poly-dispersity makes particles hard to adjust vacant spaces and
further increase of $\sigma$ introduces smaller particles filling up
vacant spaces. It is an open and interesting mathematical question if
one can rigorously prove this effect in general.

We have aslo studied the conitinuum percolation of poly-disperse systems 
in two and three dimensions and have shown that the critical area fraction
in two dimensions and the critical volume fraction in three dimensions
have opposit dependence on the poly-dispersity.
Furthermore, the critical area (volume) fraction is shown not to depend
much on the distribution function.
These results indicate that the concept of the critical area (volume) fraction
as a dimensional invariant\cite{scher} has to be applied with some caution.

\medskip

This work was supported in part by the grant from the Ministry of
Education, Science, Sports  and Culture.


\medskip
\begin{thebibliography}{99}
\bibitem{Bernal}
J. D. Bernal, Nature {\bf 183}, 141 (1959).
\label{Bernal}
\bibitem{Alder}
B. J. Alder and T. E. Wainwright, J. Chem. Phys. {\bf 27}, 208 (1957). 
\label{Alder}
\bibitem{Jaeger}
H. M. Jaeger and S. R. Nagel, Science {\bf 255}, 1523 (1992).
\label{Jaeger}
\bibitem{katoh}
R. Katoh, T. Hihara, D. L. Peng and K. Sumiyama, Appl. Phys. Lett. {\bf 82},
2688 (2003).
\label{katoh}
\bibitem{MJS1}
S. Torquato, T. M. Truskett and  P. G. Debenedetti, Phys. Rev. Lett. {\bf 84},
 2064 (2000).
\label{MJS1}
\bibitem{MJS2}
T. M. Truskett, S. Torquato and  P.G. Debenedetti, Phys. Rev. E {\bf 62},
993 (2000).
\label{MJS2}
\bibitem{okubo}
T. Okubo and T. Odagaki, J. Phys.: Condens. Matter {\bf 16}, 6651 (2004).
\label{okubo}
\bibitem{berryman}
J. G. Berrymen, Phys. Rev. A {\bf 27}, 1053 (1982).
\label{berryman}
\bibitem{Laguerre}
B. J. Gellatly and J. L. Finney, J. Non-Cryst. Solids {\bf 50}, 313 (1981). 
\label{Laguerre}
\bibitem{Ono}
G. Y. Onoda and E. G. Liniger, Phys. Rev. Lett. {\bf 22}, 2727 (1990).
\label{Ono}
\bibitem{scher}
H. Scher and R. Zallen, J. Chem. Phys. {\bf 53}, 1421 (1970).
\label{scher}
\bibitem{Bou}
D. Bouvard and F. F. Lange, Acta metall. mater. {\bf 39}, 3083 (1991).
\label{Bou}
\bibitem{Ogata}
R. Ogata, T. Odagaki and K. Okazaki, J. Phys.: Condens. Matter {\bf 17},
4531 (2005).
\label{Ogata}
\end{thebibliography}
\end{document}